\documentstyle [12pt]{l-aa}
\begin{document}
\thesaurus{11         % A&A Section 11: Extragalctic astronomy
      (11.09.2; 
       11.05.2; 
       11.16.1;
       11.19.7;
       12.03.3)}
   \title{Studies of the association of faint blue and luminous galaxies 
         using the Hitchhiker parallel camera}  
   \author{J.~B.~Jones\inst{1},   S.~P.~Driver\inst{2},  S.~Phillipps\inst{3},  
   J.~I.~Davies\inst{1},  I.~Morgan\inst{1}  \and  M.~J.~Disney\inst{1} }
   \offprints{J. B. Jones} 
\institute{
   Department of Physics and Astronomy, 
   University of Wales College of Cardiff, 
   P. O. Box 913, Cardiff, CF2 3YB, Wales, United Kingdom 
\and
   School of Physics, University of New South Wales, 
   Sydney, NSW 2052, Australia 
\and
   Astrophysics Group, Department of Physics, 
   University of Bristol, 
   Tyndall Avenue, Bristol, BS8 1TL, England, United Kingdom} 
\date{Received 11th September, 1995; Accepted 11th June, 1996}
\maketitle
\markboth{J. B. Jones et al.}{The association of faint blue and luminous 
 galaxies}
%
%  \maintitlerunninghead{The association of faint blue and luminous galaxies.}
%  \authorrunninghead{J. B. Jones et al.}
% 
\begin{abstract}
     At B magnitudes $\ga 24$ there is a well-known excess of 
galaxies (compared to standard models) which is probably due to an 
(evolving) population of sub-L$^\ast$ galaxies at moderate redshifts 
($\la 0.4$). One particular hypothesis which is hard to test 
directly via number counts or even redshift surveys is the 
possibility that the faint blue galaxies are in fact sub-galactic 
objects destined to merge by the present day to form current giant 
galaxies. If this were the case we might expect to find the faint blue 
galaxies in the vicinity of $\simeq \rm{L}^\ast$ galaxies (at redshifts 
$\simeq 0.2$ to 0.4) with which they can merge (the blue galaxies 
are already known to be weakly clustered among themselves, limiting 
the possibility for multiple mergers of small fragments). 

In the present paper we look for evidence of such clustering 
of faint blue galaxies around larger systems using candidates chosen 
photometrically from deep multicolour CCD images using the 
Hitchhiker parallel CCD camera. 
A sample of candidate L$^\ast$ galaxies expected to lie at redshifts 
$z \simeq 0.2$ to 0.4 has been selected on the basis of 
apparent magnitude (B $= 20{\fm}5$ to $22{\fm}0$) and colours 
typical of early-type spirals. 
The distribution of 152 blue galaxies having 
$23{\fm}5 < \mbox{B} < 25{\fm}0  \,,$ 
$(\mbox{B}-\mbox{R})_{_{\rm C}} < 1{\fm}2\,,$ around 13 
candidate L$^\ast$ galaxies has been determined. 
No evidence has been found for any preferential clustering of blue 
galaxies about the L$^\ast$ candidates~; the observed overdensity 
within 60~arcsec of the L$^\ast$ candidates is $-0.02\pm0.76$ per 
candidate. 
We have also looked for clustering 
between other photometrically selected samples (such as faint blue 
and faint red objects). Null results have been found in all cases, 
placing significant limits on the scenarios wherein dwarfs at medium 
redshifts are removed via mergers with larger objects. 
\end{abstract}
%
%
%
%________________________________________________________________
 
\section{Introduction}
 
      Deep imaging surveys have revealed a large population of blue 
galaxies at faint magnitudes (e.g. Tyson 1988; Lilly, Cowie \& 
Gardner 1991; Metcalfe et al. 1991, 1995; Driver et al. 1994a), 
which becomes increasingly important at fainter limits. The exact 
nature of these objects remains uncertain; for instance 
it is not clear whether 
these blue galaxies themselves have more extreme properties at 
fainter magnitudes, or whether their relative numbers increase. 
Spectroscopic surveys suggest that they are found predominantly 
at moderate redshifts ($z \simeq 0.1$ to $0.6$) 
(e.g. Colless et al. 
1990, 1993; Cowie, Songaila \& Hu 1991; Lilly, Cowie \& Gardner 1991; 
Glazebrook et al. 1995a) 
and that they tend to have strong [O{\footnotesize II}] emission 
(Koo et al. 1995). 
Since they are not predicted by standard galaxy models, 
these blue galaxies have often been cited as the strongest evidence for 
galaxy evolution with redshift (Broadhurst, Ellis \& Shanks 1988; 
Broadhurst, Ellis \& Glazebrook 1992).

     Any convincing explanation of their nature must account for 
the absence of any clear local counterpart to the faint blue 
galaxies. One possible model envisages that these objects 
have faded over time, implying luminosity evolution of either the 
entire galaxy population (e.g. Lilly, Cowie \& Gardner 1991) 
or of a subset of it (Broadhurst, Ellis \& Shanks 1988; 
Phillipps \& Driver 1995; Driver et al. 1995a). 
Alternatively, the blue galaxies may have experienced density evolution. 
Merging (Rocca-Volmerange \& Guiderdoni 1990; Broadhurst, Ellis \& 
Glazebrook 1992; Carlberg 1992) 
of the faint galaxies, either with each other or 
with more luminous objects, provides a mechanism for reducing their 
numbers by the present time (Koo 1990; Carlberg \& Charlot 1992). 

Giraud (1992) performed high resolution imaging of 
samples of blue galaxies, identifying three distinct morphological 
classes. More recently, Colless et al. (1994) have studied the 
light profiles of a sample of faint blue galaxies from the 
Colless et al. (1990, 1993) redshift survey using images taken 
in excellent seeing with the Canada-France-Hawaii Telescope. 
They found that 
faint galaxies exhibiting [O{\footnotesize II}] line emission often 
have nearby companions, suggesting interactions are important in 
activating star formation. 

Recent HST results show a large fraction (around a half) of faint galaxies to 
have irregular or peculiar morphology (e.g. Casertano et al. 1995;
Driver et al. 1995a,b). Of these
maybe one third (i.e. $\sim$ 15\% of the total population) appear to be
interacting (e.g. Driver et al. 1995b;
Glazebrook et al. 1995b). This clearly suggests a role for interactions
or mergers in the evolution of the faint blue galaxy population. Furthermore,
Burkey et al. (1994) found that $\sim$ 34\% of HST galaxies at redshifts
0.4 to 0.7 had close companions compared to $\sim$ 7\% locally, suggesting
that $\sim$ 13\% of the distant population may have disappeared by
merging. However, Woods, Fahlman \& Richer (1995) using a similar technique find
no excess pairs in the deep data.

     Various different merging models have been advocated. The faint blue 
galaxies may gradually merge with one another to form more massive objects, 
or may be accreted into massive dark haloes (Rocca-Volmerange \& 
Guiderdoni 1990). 
Cowie, Songaila \& Hu (1991) suggested that the faint 
galaxies have `parent' giant galaxies with which they have since 
merged, implying a physical (clustering) association between the 
two (see also Cowie, Hu \& Songaila 1995). 
Kauffmann, Guiderdoni \& White (1994) have fitted faint number counts using 
detailed models of hierarchical galaxy formation (in a cold dark 
matter context) in which satellite galaxy haloes merge with more 
massive dark haloes of giant galaxies. 
On the other hand, Dalcanton (1993) has argued that the conservation 
of luminosity during mergers of the blue galaxies would lead to an 
excess integrated luminosity over that observed. 
The observed thinness of the discs of spiral galaxies 
may also constrain the 
importance of mergers in the evolution of these objects (e.g. 
T\'{o}th \& Ostriker 1992). The general level of clustering among 
faint blue galaxies appears to be low (Efstathiou et al. 1991; 
Pritchet \& Infante 1992; see also Neuschaefer \& Windhorst 1995) 
which may constrain direct merger models. On the other hand, Cole 
et al. (1994) find that dwarfs and giants at moderate redshifts 
occupy the same general structures and have very similar large-scale 
clustering properties.

A model of mergers of dwarfs with giants might therefore 
be tested by measuring the small-scale clustering of faint 
blue galaxies around a sample of candidate giant galaxies; 
this is the approach we adopt here (see Jones et al. 1994). 
A preferential clustering of the faint blue galaxies around giants 
would imply that they are dwarfs at similar redshifts to the giants 
and might favour merging models over fading in accounting for the 
lack of these low luminosity systems in the nearby Universe. 
Conversely, the absence of any excess around giants might be 
interpreted as evidence against the merging of dwarfs with more 
luminous parents (see also Bernstein et al. 1994). 
Using data from the Hitchhiker camera, we study the numbers 
of these objects around photometrically selected candidate giants 
and compare them with a random distribution in order to search 
for an excess consistent with the blue galaxies being dwarfs 
associated with parent giants. 

     In Sections~2 and~3 we describe the observational data 
and the data reduction methods. In Section 4 the image detection 
process is detailed together with the techniques used for aperture 
photometry. The definition of samples of faint blue galaxies and 
candidate luminous galaxies are discussed in Section 5, and the 
association of the blue galaxies with the giants is determined. 
We model the distribution of random samples of faint images to 
account for the effects of the limited areas of data frames and 
to demonstrate that the light of bright images does not mask 
faint images to effect significantly the statistics of fainter 
objects. Section~6 considers the association between other 
photometrically selected samples. Section~7 presents a detailed 
investigation of the errors in the analysis. 
Finally, the implications of the results for the evolutionary history 
of the galaxy population are briefly considered.

\section{Hitchhiker camera observations}

     The data consist of deep B- and R-band images of four fields obtained 
with the Hitchhiker parallel CCD camera on the 4.2m William Herschel 
Telescope on La Palma. The camera, described in Driver (1994), performs 
imaging in an off-axis field, 7~arcmin from the optical axis, 
while the telescope is used for its normal spectroscopic 
programmes. Using a dichroic beam splitter, data are 
recorded from the same field in two colours simultaneously. 

The data used here were collected over several observing periods in 
1991 and in 1993. They are summarised in Table~1. Example R band 
data are presented in Fig.~1, where the candidate L$^\ast$ and faint 
blue galaxies of Section~5.1 are marked. 
The 1991 data are described in detail by Driver (1994) 
(and one field by Driver et al. 1994a). 
Total integration times on the fields ranged between 32 and 
125~minutes through each filter. 
Individual exposures were typically of 5--10 minutes duration. 
The fields used were all at moderate or high galactic latitude. 
Seeing was around 1.5~arcsec (full-width at half-maximum) in the R band; 
image size measurements made with the {\footnotesize KAPPA} software package, 
supplied by the Starlink project, are presented 
in Table~1. Reduced frames were all about $5' \times 3'$ in extent; 
this means that clustering 
on scales up to $\sim 1\,$Mpc can be investigated at the distances at 
which we expect to see L$^\ast$ galaxies. 
Note that if merging takes a few (say 5) dynamical times, only neighbours 
within $\sim\frac{1}{3}\,$Mpc of a galaxy with rotation/dispersion 
velocity $\sim 300\,\mbox{kms}^{-1}$ could be expected to merge in 
the $\sim 5\,$Gyr since $z \sim 0.4\,.$ 
Typically 200--300 galaxies are detected in each field.

%
%   TABLE 1
% 
% \setcounter{table}{0}
%
% \begin{table*}
%\label{t3}
% \label{1}
% \begin{center}
% \caption{Observational data.}
% \mbox{ } \\[0mm]
% \begin{tabular}{cclcccc}
% R.A.  & Dec.  & 
%    Date  & \multicolumn{2}{c}{ Integration time (min.) } &
%    R band seeing & Area of reduced   \\
% {\scriptsize (1950.0)} & {\scriptsize (1950.0)} & & \hspace{2mm} B band 
%    \hspace{1mm} & \hspace{3mm} R band \hspace{1mm} &
%    (FWHM) & data frame \\[3mm]
% $12^{^h} 29.5^{^m}$  &  $+26^{^o} 26'$  &  1991 Feb 15 & 120 & 120 & 
%    1.6 \mbox{arcsec} & $5.5' \times 3.0'$ \\[3mm]
%  11~ 40.1~ & +19~ 49~~ & 1991 Feb 12        & 75 & 75 & 1.3~\hspace{8mm} & 
%     $5.5 \times 3.0$ \\[3mm]
%  13~ 37.8~ & +11~ 28~~ & 1993 Feb 28, & 40 & 60 & 1.6~\hspace{8mm} & 
%     $5.0 \times 3.3$ \\
%            &           & \hspace{9mm} Mar 2 & & &     & \\[3mm]  
%  15~ 47.6~ & +21~ 35~~ & 1993 May 22        & 32 & 32 & 1.6~\hspace{8mm} & 
%     $5.3 \times 3.1$  \\[8mm]
% \end{tabular}
% \end{center}
% \mbox{ } \\[5mm]
% \end{table*}

\section{Data reduction} 

     The parallel mode of operation of the 
Hitchhiker camera, without any 
control of the pointing of the telescope, requires that the data 
be reduced in a slightly non-standard manner. 
These methods are described in full by Driver (1994) and 
summarised by Driver et al. (1994a). 
Firstly, bias signal subtraction was performed by 
removing a mean bias level calculated from the CCD bias strips 
of each raw data frame. The validity of this approach has been 
verified by testing the intensity of the bias signal across the 
whole frame; no significant structure has been found. 
The data frames suffer from an essentially 
circularly symmetric vignetting pattern (caused by 
instrumental optical components). Because this varies with 
the focussing of the camera lenses, and because it is not always 
possible to obtain nightly flatfielding data, it is necessary 
to correct for this vignetting at the outset of data reduction. 
For the 1991 observations 
the vignetting pattern was modelled by measuring the sky background 
at a set of points on a rectangular grid, performing two-dimensional 
bicubic spline interpolation between the sampling points. A relatively 
coarse grid was chosen, 5 by 7 sampling points in extent, in order to 
lessen the possibility of removing any extended astronomical structures. 
The frames were then corrected for vignetting by division by the model. 
For the 1993 data a median filtering technique using a 45~arcsec wide filter 
box was employed. Flatfielding 
was accomplished for the 1991 data using a superflat constructed 
from over 100 long exposures of the night sky. For the less extensive 
1993 observations the superflats were generated by coadding a number 
of twilight sky frames. 
The individual flatfield frames were vignetting-corrected 
before coaddition to produce the superflats; in this application the 
flatfielding process corrects only for the pixel-to-pixel efficiency 
variations and not for large-scale effects. 

Cosmic ray detections were removed from the individual dark sky frames 
before alignment and coaddition. Candidate cosmic ray detections were 
identified as localised peaks rising higher than seven standard 
deviations above the sky background. The intensity profiles of the 
peaks were determined, with those steeper than a typical seeing disc 
being labelled as a genuine cosmic ray event. The pixel intensities 
in each of these detections were set to the median in surrounding 
pixels. 

Finally the coadded images were cleaned by performing an additional 
median filter sky subtraction (with a 40~arcsec wide 
square box). 
Remaining spurious, low-intensity artifacts (due to dust particles on the 
instrumental optical surfaces) were removed via inverse unsharp masking 
({\it cf.} Driver et al. 1994a). 
The photometric calibration of 
the data was accomplished using the results of pointed Hitchhiker 
observations of standard stars in February 1991 and May 1993, 
corrected to the date of observation using the nightly extinction 
coefficients from the Carlsberg Automated Meridian Transit Circle 
Telescope on La Palma. Magnitudes are expressed on the Cousins 
BVR$_{_{\rm C}}$I$_{_{\rm C}}$ system (Cousins 1976; Bessell 1979), calibrated 
with Landolt (1983) photometric standards.  
The accuracy of the magnitude scale has been shown 
to be $0\fm1$ in each filter (Driver 
et al. 1994b). Colour indices are, however, more 
accurate because of the simultaneous recording of data through the two 
filters.

\section{Image detection and photometry }

     Image detection was accomplished with a connected-pixel algorithm, 
using the {\footnotesize IMAGES} program of the {\footnotesize RGASP} 
galaxy photometry software package (Cawson 1983). For classification 
as a genuine image, a group of ten or more adjacent pixels 
($\simeq 1\,$arcsec diameter) had to have R band intensities above a threshold 
of $1.5\sigma_p$ above the sky background (where $\sigma_p$  
represents the pixel-to-pixel standard deviation of the sky background). 
This minimum number of pixels corresponds to the expected minimum area 
of any reliably detected image due to the size of the seeing disc; 
given the image scale of 0.3~arcsec/pixel, this corresponds to the 
area within the half-maximum intensity isophote of a star 
under good seeing. 
The $1.5\sigma_p$ threshold was chosen to 
exclude a significant chance contribution of background pixels 
to the area of detected images (Driver 1994). A conservative 
estimate of the background standard deviation was taken, based 
on adopting the largest value from either: the measured background 
variation;
a theoretical prediction of the noise in the background assuming 
Poisson statistics; or a value of $0.3\%$ of the background level 
(based on the expectation of a $0.3\%$ limiting accuracy of the 
flatfielding process on large scales). 
In practice, the Poissonian prediction of the background standard 
deviation was adopted for all four fields. 
To safeguard against 
spurious detections of random groups of sky background pixels, 
a signal-to-noise ratio test was used to reject low confidence detections. 
Monte-Carlo tests were performed on simulated data frames constructed 
using Poissonian noise distributions in order to assess the number of false 
detections retained after imposing different signal-to-noise ratio limits. 
Further simulations were carried out using twilight sky frames subjected 
to the same data reduction procedure as the night sky data. On the 
basis of these tests, a signal-to-noise ratio limit of 6.0 for the isophotal 
data was adopted. This value was found to give as few as one or two false 
detections per frame for the random noise simulations. 

     Once R band catalogues of images on the data frames had been compiled, 
magnitudes were determined using (variable) aperture photometry. 
In contrast to isophotal photometry, this technique should 
measure all the flux from a detected object when used with a 
large enough aperture 
size. Ideally the radius of the aperture should be chosen for a 
particular galaxy to include essentially all the signal 
from the galaxy, but not so large that it includes unnecessary noise 
from the sky background or nearby sources. Tyson (1988) noted 
that, as expected, isophotal magnitudes are close to the total magnitudes for 
bright objects, while Metcalfe et al. (1991) showed that 
aperture photometry using Kron radii (Kron, 1980) results in fixed 
aperture sizes for faint objects (effectively the Kron radius for 
a star). We therefore chose 
a variable circular aperture radius $r_{ap}$ computed from the isophotal 
radius $r_{iso}$ as, 
\begin{equation}
   {r_{ap}}^n  =  {r_{iso}}^n  +  {r_{min}}^n  \hspace{5mm} ,
\end{equation}
where $r_{min}$ and $n$ are constants. $r_{iso}$ was calculated 
from the number of pixels having intensities above the threshold 
of the {\footnotesize RGASP} detection 
process, being the radius of a circular region containing that number. 
$r_{min}$ is set to 3~arcsec, 
about twice the typical seeing width (and in keeping with Metcalfe {et al.,} 
1991, and Lilly, Cowie and Gardner, 1991). The aperture radius 
therefore reduces to 3~arcsec for the faintest objects while approaching 
the isophotal radius for the brightest. The optimal value 
of the exponent $n$ was selected on the basis of simulations of 
the measurement of images of face-on L$^\ast$ exponential disc galaxies; 
being circular and lacking bulge or nuclear components, these provide 
the most extended and flattest profiles among the conventional galaxy 
population. 
The values of $r_{ap}$ for different values of $n$ were calculated 
for different magnitudes and compared with the isophotal radius, the 
Kron radius and the radius containing $90\%$ of the light. 
Using an exponent of $n=1.5,$ $r_{ap}$ was found to be close to the 
$90\%$ light radius over a wide range of total magnitudes, even at 
the faintest limits, and close to 2.5 Kron radii; we therefore chose to 
adopt $n=1.5\,.$ Figure~2 shows the dependence of the total detected 
magnitude within a circular aperture for different aperture radii for 
the case of face-on, exponential light profile, L$^\ast$ galaxies. 
The various curves in the figure represent different methods for defining the 
aperture radius. 

     A local measurement of the sky background surface brightness 
was used to remove the sky background contribution from the total 
signal within the aperture for each image. This was defined as the 
median of the pixel intensities in a 15-pixel wide circular 
annulus centred on the image, 
having an inner radius of $2.0\,r_{ap}\,,$ excluding pixels which 
themselves lay within the inner radius of the equivalent annuli used 
to determine the background level around other images. In this way, 
an estimate of the background level was obtained which was essentially 
free of the contributions of detected images. 

     To avoid problems associated with incomplete data at the edges of 
the frames, only object images whose centres lie further than 30 pixels 
from the edges are considered. Fuller details of the image detection and 
photometric techniques are presented by Driver (1994). 

     The determination of the observed properties of galaxies is 
complicated by the overlapping of images through chance alignments. 
The reliable decoupling of blended images is a difficult process, 
complicated by factors such as the uncertainty in deciding how to 
assign the signal in the merged regions between the images, 
and the dependence of the efficiency of the process on the brightness 
of the image. For this analysis, if the isophotes of the two objects 
overlapped we 
simply considered the system as merged and counted it as a single image. 
The effects of the overlapping of images in detecting and 
parameterising faint galaxies in the vicinity of brighter ones are 
discussed in Section~5.3, where it is shown that overlapping 
images do not significantly affect the clustering statistics of 
interest here. 

      Once these principles had been used to provide R-band magnitudes 
for each detected image, B magnitudes were computed using the same 
(R band) image catalogue and the R band apertures. 
This method ensures that each image is treated identically in each of 
the two wavebands in an effort to minimise photometric errors in 
the colour index. 

      The photometric results for all four fields are displayed as 
a colour--magnitude diagram in Figure~3, showing all images, both 
stars and galaxies. The broad distribution is similar to that found 
by other authors (e.g. Tyson, 1988, and Metcalfe et al., 
1991). 
The faint blue excess, however, is encountered about one magnitude 
brighter at any given (B--R)$_{_{\rm C}}$ colour than in many other 
studies. 
This effect is found 
to be pronounced among the 1993 field data, but not those 
from 1991. That this is not a calibration problem affecting the 
1993 data is confirmed by 
an inspection of the (B--R)$_{_{\rm C}}$ against (V--I)$_{_{\rm C}}$ 
colour--colour diagram for 
brighter ($R < 21\fm0$) images; a majority of the images, which at 
these magnitudes are expected to contain a significant fraction of stars 
(50--60\%), conform closely (within $\pm 0.15\,\mbox{mag}$) 
to standard (Bessell 1979, and Bell \& Gustafsson 
1979, 1989) stellar loci. The problem therefore affects only fainter 
images  -- if indeed it is a problem rather than some statistical fluke. 
While it is expected that random photometric errors will be 
greater for the 1993 
observations due to their shorter integration times, 
the origin of the difference remains unclear. 
However, for the purposes of the present study, where colours and 
magnitudes need to be measured only sufficiently accurately to enable 
a broad classification, the excess blue tail to the colour distribution 
at faint magnitudes is unimportant.

\section{The clustering of faint blue galaxies around L$^\ast$ galaxies}

     It should be possible to test the hypothesis that the faint 
blue galaxies are in reality dwarfs associated with giant galaxies 
if a sample of candidate giants can be identified on the basis 
of their observed photometric properties. We can then test the 
clustering of the blue galaxies around them. 
Magnitude-limited samples tend to select galaxies having luminosities 
close to L$^\ast$ for galaxies with properties comparable with the 
local population (e.g. Schechter 1976). 
Even for their dwarf-rich model of galaxy populations, 
Driver et al. (1994a) 
have shown that magnitude-limited samples of galaxies are likely 
to be dominated by giants (with 
intrinsic luminosities close to L$^\ast$ of the Schechter luminosity 
function) for B $< 22^{^m}$;
only at fainter magnitudes do 
they predict that dwarfs become increasingly important. 
A more conservative field galaxy luminosity function, without a 
sharp turn-up in the numbers of dwarfs, would predict giant 
domination to even fainter magnitudes. That galaxies in the 
B $= 20^{^m}$ to $22^{^m}$ magnitude range are dominated by L$^\ast$ 
objects is confirmed from the results of redshifts surveys (e.g.
Broadhurst, Ellis \& Shanks 1988). 
Expecting samples of galaxies 
brighter than (at least) B $= 22^{^m}$ to be rich in 
giant galaxies, we select candidate L$^\ast$ galaxies 
on the basis of magnitude and colour. 

     A simple measure of the clustering of a sample of objects about 
a central point is the excess over random statistics of the objects 
within a particular distance of the centre 
(e.g. Yee \& Green 1987;  Longair \& Seldner 1979; 
Arag\'{o}n-Salamanca et al. 1993). 
Such a method is consistent with the use of the two-point angular 
correlation function (Phillipps \& Shanks 1987a,b). 
Note that here we are not concerned about detecting an average excess 
over some overall random background; rather we are interested in the 
distribution of one specific set of galaxies about another specific set. 
We therefore do not need to consider the intricacies of data--data, 
data--random or random--random pairs (e.g. Landy \& Szalay 1993),
but simply calculate the numbers of 
faint blue galaxies in concentric annuli of constant thickness centred 
on the candidate L$^\ast$ objects and compare these with the corresponding 
results for randomly distributed points.

\subsection{Selecting samples of galaxies}

     We choose to select a sample of candidate L$^\ast$ galaxies 
using a B  magnitude range of 1.5~mag. extending to the 
B $= 22{\fm}0$ 
approximate limit of giant domination of the observed galaxy population. 
We further constrain the sample by imposing limits in (B$-$R)$_{_{\rm C}}\,.$
The no-evolution model of galaxy colours used in Driver et al. (1994a) 
provides the mean (B$-$R)$_{_{\rm C}}$ as a function of B magnitude for Sa  
giants ({\it cf.} Coleman, Wu \& Weedman 1980). 
We set a red limit for the candidate L$^\ast$ sample at 
$0{\fm}2$ redder than this locus to allow for photometric 
errors and to take some account of a distribution of galaxy 
properties, e.g. from E to Sc. If present, evolutionary 
effects would produce galaxies bluer than these colours. We define a 
locus in the (B$-$R)$_{_{\rm C}}$ -- B plane accounting for evolution of the 
stellar populations by introducing the blueing 
effects of Bruzual's (1983) $\mu = 0.5$ models of spiral bulges 
and elliptical galaxies on the Driver et al. Sa giant model. 
A blue limit $0{\fm}2$ bluer than this evolution model 
is used for the candidate L$^\ast$ selection. The resultant 
sample contained 17 photometrically-selected images. 
For an ${\Omega}_0 = 1, {\lambda}_0 = 0$ cosmological model, 
these L$^\ast$ galaxies are expected to lie at redshifts of 
$z \simeq 0.2$ to $0.4$ ({\it c.f.} Koo \& Kron 1992). 
Throughout this paper a Hubble Constant 
of $H_{_0} = 50\,\mbox{kms}^{^{-1}} \mbox{Mpc}^{^{-1}}$ is used. 

     Star--galaxy classification was attempted in order to reject star 
images from the sample of candidate $L^{\ast}$ galaxies. 
Stars are likely to form a significant fraction of objects having the 
magnitudes of the $L^{\ast}$ candidate sample and the images are 
sufficiently bright that image classification can be attempted. 
Following Jones et al. (1991), we used plots of the 
image central intensity against R magnitude, and of the area above the 
detection isophotal threshold against R magnitude for each data frame. 
Stellar loci were identified for each plot. The images of interest 
were classified according to their displacement from the appropriate 
stellar locus. 
The images were labelled as being stars, galaxies or 
having uncertain classifications, the process being performed 
independently for the central intensity and image area 
graphs. 
An overall classification was achieved through a comparison of 
the results of the intensity and area methods, and by visual 
inspection to reject merged or confused images. An object was 
regarded as a suitable L$^\ast$ galaxy if it appeared visually to 
be a single image and if it had received a galaxy classification 
under the automated tests, either through two unambiguous galaxy 
classifications or one galaxy and one uncertain result. 
Of the 17 objects in the original sample, 13 passed the star/galaxy 
tests. These 13 images formed the sample of candidate L$^\ast$ galaxies  
for the present study. 

     The sample of faint blue galaxies was selected using an apparent 
magnitude range of B $= 23{\fm}5$ to $25{\fm}0\,;$ these limits are 
$3{\fm}0$ fainter than the equivalent ones for the candidate L$^\ast$ objects. 
The adopted colour limits were (B$-$R)$_{_{\rm C}} = -0{\fm}4$ 
to $+1{\fm}2\;.$ 
An additional criterion was imposed that the galaxies (of all four 
fields) lay within the predicted selection limit of the 1991 
February 15 field (see Table~1); it is to be expected that 
photometric results outside this limit are unreliable due to the low 
signals involved. This limit is shown in the colour--magnitude 
diagram in Figure~3.

      No star/galaxy classification was attempted for the faint blue 
sample; at these faint magnitudes and blue colours the sample of images 
is dominated by galaxies, as is evident from a comparison of the 
star count predictions of Bahcall \& Soneira (1980) with standard 
galaxy number  counts. 
Indeed, at these faint magnitudes it becomes very difficult to distinguish 
galaxies from stars given the small image sizes compared with 
the seeing discs. 
To illustrate this point more fully, the 
numbers of stars expected in the faint blue galaxy samples in the four
fields were computed by modelling star number counts. Using a program 
written and provided by Dr. G. Gilmore (briefly discussed in Gilmore 1984), 
star densities were computed 
across the (B--V) -- V colour--magnitude diagram for each field 
by integrating the stellar populations along the sight. The three-component 
Gilmore--Reid--Wyse model of the Galaxy was adopted 
(Gilmore, Wyse \& Kuijken 1989; Gilmore, King \& van der Kruit 1990). 
Converting to the (B$-$R)$_{_{\rm C}}$ -- B colour--magnitude diagram and integrating 
the predicted star densities over magnitude and colour provides estimates 
of the numbers of stars in the faint blue galaxy samples in each 
of the four fields. In all cases these are small, between 1 and 4 stars, 
as a result of the colour index limits of the sample being bluer than the 
majority of the main sequence stars of the (old) Galactic halo and thick 
disc. Due to incompleteness of the samples and to photometric errors, 
it is expected that the numbers of stars {\em observed} in the faint blue 
sample in each field will be smaller than the estimated numbers of stars 
present. We therefore choose to express the star contamination as a 
fraction of the total number of images present, taking the total 
numbers of images from the deep observations of Metcalfe et al. 
(1995) which are complete in all the regions of the (B$-$R)$_{_{\rm C}}$ -- B 
colour--magnitude
diagram of the photometrically selected samples used here. We predict 
a star contamination of the faint blue galaxy samples of between 
$1.0\,\%$ and $2.7\,\%$ in the four fields; these results are 
presented in full in Table~5. The presence of such small 
numbers of stars will not significantly affect 
conclusions about the clustering of the faint blue galaxies. 

       The faint blue galaxy sample contains 152 objects over all 
four fields. Because of the different K-corrections, 
the magnitude limits of the L$^\ast$ and blue galaxy samples are 
displaced typically by $3{\fm}6$ in absolute B magnitude. 
The luminosities of the faint blue galaxies, if at the same 
distances as the L$^\ast$ objects, would be in the range 
$\mbox{L}_B = 0.01\mbox{L}_{\ast}$ to $0.1\mbox{L}_{\ast}\,,$ 
typically $\mbox{L}_B = 0.04\mbox{L}_{\ast}\,.$ 
Figure~3 shows the sample regions in the colour -- magnitude 
diagram. 
Although the adopted selection criteria should produce a well-defined 
sample of blue galaxies, incompleteness in the R band catalogue for the 
less deep fields may bias the samples against the most extreme blue 
galaxies, possibly reducing the sensitivity of the results to the most 
extreme colour changes induced by galaxy interactions.

\subsection{The statistics of the separations between the blue and L$^\ast$ 
  galaxies.}

      The separations between each of the candidate L$^\ast$ and blue 
galaxies were computed for the four fields from their R band centroid 
coordinates. The R band observations tend to be deeper than the 
B band data on account of the greater efficiency 
of the Hitchhiker camera at red wavelengths than blue; the R band 
positional data were therefore used in preference to the B band.  
A total of 525 separations were obtained. 
Figure~4 presents a histogram of the separations between the 
faint blue and candidate L$^\ast$ galaxies summed over all four 
data frames. 
The deviation of the observed separation distribution from an ideal 
linear relation is due to the finite area of the data frames. 
An assessment of any excess density of blue galaxies around L$^\ast$ 
objects demands that the histogram of separations for random 
distributions of galaxies is known. 

     Monte Carlo simulations were used to model randomly 
distributed faint blue galaxies across the four data frames. 
This approach enables the effect of the finite areas of the 
data frames to be accounted for in detail. 
In order to overcome the statistical errors associated with 
small samples, 
$100\,000$ faint galaxies were distributed across each frame and the 
separations between these and the 13 observed L$^\ast$ galaxies 
were computed. The distributions of separations for each of the 
four frames were normalised and added according to the number 
of separations from the observational data for each frame. 
The resultant distribution may be compared directly with the equivalent 
observational results; both histograms are shown in Figure~4. 

      To investigate whether there is an overdensity of faint blue 
galaxies in the vicinity of L$^\ast$ candidates, the number of 
separations between the two samples observed 
in the 10 to 60~arcsec range relative to 
the total number of separations were compared with the Monte 
Carlo simulations. The results are presented in Table~2. 
The observed results are clearly consistent with no observed overdensity of 
blue galaxies around L$^\ast$ candidates on scales smaller than 
1~arcmin compared with the entire $0 - 5\,$arcmin range.

%
%   TABLE 2
% 
% \setcounter{table}{1}
%
% \begin{table*}
% \label{2}
% \begin{center}
% \caption{Observational results for the association of faint blue galaxies 
%  with candidate L$^\ast$ galaxies. }
% \mbox{ } \\[0mm]
% \begin{tabular}{ccc} 
% Separation range & Observed number of galaxy-- & Predicted number 
%    of separations \\
%  (in arcsec) & galaxy separations & for random distributions of 
%   blue galaxies \\[3mm]
%     all   & 525 & $525.00\dagger$ \\[3mm]
%  10 -- 60 & $92 \pm 10$ & ~92.26 \\[3mm]
%  15 -- 60 & $88 \pm 9$  & ~88.40 \\[8mm]
% \end{tabular}
% \end{center}
% $\dagger$ Set by the normalisation to the number of observational results. 
% \mbox{ } \\[5mm]
% \end{table*}

\subsection{The detection of faint images in the vicinity of brighter galaxies.}

      A potential problem which complicates the interpretation of the 
results of Table~2 is that of a failure to detect faint galaxy images 
in the close vicinity of brighter galaxies (cf. Turner et al. 1993). 
At small separations 
galaxy images might become merged at the limiting detection isophote 
and the pixels of the fainter image might be included with those of 
the brighter object during the compilation of the image catalogue. 
A selective loss of faint galaxy images at small separations could conceal the
presence of a genuine excess of faint blue galaxies around the 
candidate L$^\ast$ objects. 

      The 13 galaxies of the L$^\ast$  sample have image areas above 
the detection threshold corresponding to mean radii in the range 2.8 to 
5.3~arcsec. The images of the 152 blue 
galaxies have mean total radii typically in the range 0.7 to 1.7~arcsec. 
It is therefore to be expected that galaxy--galaxy separations of 
10~arcsec and greater will not be significantly affected by the merging 
of images. 
At a typical redshift of $z = 0.3$ an apparent angular separation of 
10~arcsec corresponds to a transverse physical separation of 55~kpc 
(for $H_0 = 50\,\mbox{kms}^{-1}\mbox{Mpc}^{-1},$ $q_0 = 0.5$ and zero
cosmological constant).

      To test this in greater detail, simulations were performed of the 
detection of faint images in the vicinity of example candidate L$^\ast$ 
images. The May 1993 R band data frame was selected for the study, being 
the least deep of the available R band observations. 
The frame has 4 candidate L$^\ast$ galaxies from the sample of Section~5.1. 
They have magnitudes in the range $R = 18\fm8$ to $20\fm3\,.$ 
Faint blue galaxies were represented by circularly-symmetric 
gaussian light profiles having full-widths at half-maximum intensity 
equal to the measured seeing. The central intensities were selected 
to give a total magnitude of $R = 23\fm5\,,$ typical of the faint 
blue sample. The blue galaxies were added to the observed R band 
data frame, one at a time, and the image detection process of Section~4 
applied to the frame. The image catalogue was inspected to determine 
whether the artificial blue galaxy had been detected as a distinct image, 
whether it was merged with the  L$^\ast$ galaxy, or whether it was 
merged with another nearby galaxy. Faint blue galaxies were placed 
at distances of 5.0, 7.5, 10.0, 12.5 and 15.0~arcsec from the centroid 
of the L$^\ast$ candidate, at each of 8 positions for each separation. 
A total of 160 simulated images were used. 

      Table~3 presents the results of the simulations. While merging 
of the artificial faint galaxy with the L$^\ast$ candidate is a major 
problem for separations smaller than 10~arcsec, it does not significantly 
affect separations greater than 10~arcsec. The results of Table~2 
for the interval 10 to 60~arcsec are therefore unaffected by image 
blending and our null result remains. 
Table~3 does show that merging of the blue galaxy with a 
third image does occur. It has been assumed that the distribution 
of this general background of galaxies with which some of the faint blue 
galaxy images merge is uniform across the frame, and therefore affects 
clustering statistics equally on all scales.

%
%   TABLE 3
%
% \setcounter{table}{2}
%
% \begin{table*}
% \label{3}
% \begin{center}
% \caption{Statistics for the detection of faint galaxy images in the 
% vicinity of four candidate L$^\ast$ galaxies. }
% \mbox{ } \\[0mm]
% \begin{tabular}{ccccc} 
% Separation between & Number of  & Number of times &
%    Number of times & Number of times faint \\
%
% faint galaxy and L$^\ast$  & simulations & faint and L$^\ast$ & faint galaxy 
%  was & galaxy was merged  \\
%
% candidate (arcsec)  & considered & galaxies merged & a distinct image & 
%  with a third image  \\[3mm]
%
%  ~5.0 & 32 & 32 & ~0 & ~0 \\[3mm]
%  ~7.5 & 32 & 30 & ~2 & ~0 \\[3mm]
%  10.0 & 32 & ~5 & 18 & ~9 \\[3mm]
%  12.5 & 32 & ~0 & 20 & 12 \\[3mm]
%  15.0 & 32 & ~0 & 22 & 10 \\[8mm]
% \end{tabular}
% \end{center}
% \mbox{ } \\[5mm]
% \end{table*}

\section{The associations between other samples of galaxies.} 
 
       The study of any possible association between faint blue and 
L$^\ast$ galaxies has been extended to a consideration of other samples 
of galaxies. A colour index test was applied to the brighter galaxies 
in order to select the L$^\ast$ candidates. It is of interest to 
investigate whether a relaxation of the selection criteria to include 
brighter galaxies of all colours affects the conclusions. Equally, 
any possible clustering between the brighter ($20{\fm}5 \leq \mbox{B} 
\leq 22{\fm}0$) galaxies and all faint 
($23{\fm}5 \leq \mbox{B} \leq 25{\fm}0$) galaxies, the faint blue with faint 
red galaxies, and between the faint blue galaxies and themselves 
have been investigated. 

        The selection criteria for the samples of galaxies are summarised 
in Table~4. The brighter images were subjected to a star/galaxy 
classification as described in Section~5.1 for the L$^\ast$ candidates. 
This reduced the number of images from 38 to 21. 
As in the case of the faint blue galaxies, no classification was 
attempted for the faint and faint red samples. 
Unlike the case of the faint blue samples, however, the samples will 
include modest numbers of main sequence stars of the Galactic halo and 
thick disc. 
Following the method of Section~5.1, star number densities were 
predicted for each galaxy sample of each of the four fields and 
image number densities were taken from Metcalfe et al. (1995). 
The resulting fractional star contaminations are presented in Table~5. 
The effect of stars is in general small, although the fourth field, 
lying at intermediate galactic latitude, does experience some significant 
contamination to the red galaxies sample.

       Galaxy--galaxy separations were calculated between the members 
of different samples, as presented in Table~6. The table lists the total 
number of galaxy--galaxy separations and the number of separations in the 
range 10 to 60~arcsec. 

       Detailed Monte-Carlo simulations were performed for each pair of 
samples to predict the number of separations in the 10 to 60~arcsec
range, using large numbers of random points to reduce numerical noise, 
normalising the results to the observed total number of separations in 
each frame. The bright -- faint blue galaxies simulation used the observed 
bright galaxy data, but $100\,000$ simulated random faint blue images 
for each frame. Two simulations were performed for the faint blue -- faint 
red samples: one (case~(a) in Table~6) used $20\,000$ random blue galaxies
and the observed red on each frame; the other (case~(b)) used the 
observed blue and $20\,000$ random red galaxies on each frame. The bright -- 
faint galaxy study again used the observed bright galaxy data, 
with $100\,000$ simulated random faint images on each frame. 
The simulation of the association of faint blue galaxies with themselves 
was firstly (case~(a)) performed using the observed blue galaxies and 
$20\,000$ random points on each data frame, then again (case~(b)) with 
a single random sample of 2000 galaxies per frame computing the 
separations internal to the random sample. 
All simulation results were normalised to the observed total number of 
separations.

\section{An investigation of the errors in the results}

      To assess whether there exists any observed excess 
or deficiency in the number of galaxies of one type in the vicinity 
of galaxies of another, it is necessary to understand the error in the 
number of galaxy--galaxy separations in the 10 to 60~arcsec range. 
A na\"{i}ve estimate of the error in the number of separations 
might be as the square root of the number.
However, individual separation results will be highly correlated 
and such a simplistic approach may provide an incorrect error estimate, 
leading to inappropriate statistical conclusions. 

      Detailed numerical simulations have been performed of the 
statistics of galaxy distributions. Randomly distributed galaxies 
were placed in four simulated data frames corresponding to the samples 
of Table~4, having the numbers of the observed data. The galaxy--galaxy 
separations were determined and the number in the 10 to 60~arcsec range 
calculated. This process was performed repeatedly a total of 200 
times for each pair of galaxy samples. The statistics of the 
number of separations in the 10 to 60~arcsec range provided estimates 
of the errors in the individual results of Tables~2 and~6; these are 
the errors quoted in these tables. 

      The Monte Carlo estimates of the error are larger than a 
simplistic $\sqrt{n}$ estimate, particularly for the bright--faint, 
the faint blue--faint red, and especially the faint blue--faint blue 
cases. Indeed had the na\"{i}ve estimate been used, it would have 
given a false anticorrelation between faint blue galaxies 
at a 2.6-sigma confidence level. 

      The results obtained using the detailed error assessments are 
presented in Table~7.

\section{Discussion}

     The results presented in Table~7 show no statistically significant 
evidence of any preferential clustering of galaxies of one type about 
those of another for any of the pairs of samples investigated. 
In particular, there is no evidence of any association of the faint blue 
galaxies about candidate luminous galaxies. 

      The candidate L$^\ast$ -- faint blue galaxy study found 
$92\,\pm\,10$ galaxy -- galaxy separations in the 10 to 60~arcsec
range, using 13 L$^\ast$ candidates. The predicted number for a 
random distribution of blue galaxies was $92.26\,.$ This gives 
an excess number of separations of $-0.26 \pm 10$ in this range, 
equivalent to an overdensity of $-0.02 \pm 0.76$ blue galaxies 
per L$^\ast$ candidate. 

      L$^\ast$ galaxies having B $= 20{\fm}5$ to $22{\fm}0$ 
would lie at redshifts of $z = 0.23$ to 0.36, adopting 
$M_B^\ast = -21\fm0\,,$ K-corrections appropriate to Sab 
galaxies (Driver et al. 1994), 
$H_0 = 50\,\mbox{kms}^{-1}\mbox{Mpc}^{-1},$ 
$q_0 = 0.5$ and zero cosmological constant. 
At a typical redshift of $z = 0.30\,,$ a 60~arcsec angular radius 
about the  L$^\ast$ galaxy corresponds to a physical radius of 330~kpc 
(the expected limit for potential merging victims, as discussed in 
Section~2). Thus a $2.5\,$sigma upper limit of 1.9 blue galaxies 
within this region corresponds to a projected mean excess density of 
5.6 Mpc$^{-2}.$ 

      It is informative to compare this limit with the predicted 
number of dwarf galaxies of all colours within the 330~kpc radius, 
having similar magnitudes to the blue galaxies, 
which would be given by the local galaxy population extrapolated to 
$z = 0.30\,.$ The B $= 23{\fm}5$ to $25{\fm}0$ apparent magnitude limits 
for the faint blue galaxy sample corresponds to 
$M_B = -18\fm3$ to $-16\fm8$ at $z = 0.30$ for K-corrections 
typical of dwarf irregulars, or $-19\fm2$ to $-17\fm7$ 
if they are dwarf ellipticals. In either case, using the Efstathiou, 
Ellis \& Peterson (1988) local parameterisation of 
the Schechter luminosity function (with a faint end slope 
$\alpha = -1.07$), the number density of 
conventional (having the properties of the local population) 
dwarf galaxies of all colours passing the 
apparent magnitude selection test of the faint blue galaxies is predicted 
to be $\simeq 0.008 \mbox{Mpc}^{-3}$ in the absence of evolution 
after applying a $(1+z)^3$ volume scaling. 
A steeper faint end of the galaxy luminosity function, with 
$\alpha = -1.25\,,$ would increase this by a factor of $\simeq 2\,,$ as would a
rescaling of the luminosity function to fit the galaxy counts at $B \simeq 18$
(see e.g. Metcalfe et al. 1995; Glazebrook et al. 1995c).

      From Phillipps (1985a,b) and Phillipps \& Shanks (1987a) 
we see that the expected projected excess number density of galaxies 
within an angular radius corresponding to a projected physical separation 
$s$ from a galaxy 
is related to the amplitude of the correlation function $r_0$ by 
\[
   E(<s) \;\; = \;\; \frac{2}{3-\gamma} \; G(\gamma) \; r_0^\gamma \; 
    s^{1-\gamma} \; \phi_{int}    \hspace{4mm} , 
\]
where $\gamma$ is the index of the spatial two-point correlation function 
$\xi(r) = (r/r_0)^{-\gamma}\,,$ 
$G(\gamma)$ is a constant ($=3.6791$ for $\gamma=1.8$), 
and $\phi_{int}$ is the integral of the galaxy luminosity function between 
the two absolute magnitude limits of the sample. 
Assuming a number density of faint blue galaxies equal to that of 
a conventional (local, no evolution, with only density scaling) galaxy 
population at $z \simeq 0.3\,,$ 
the $2.5\sigma$ upper limit on the clustering amplitude is then 
$r_0 < 9\,\mbox{Mpc}$ for a standard flat luminosity function or 
$r_0 < 5\,\mbox{Mpc}$ for a slightly steeper one or one with a higher
normalisation. 
If the faint blue galaxies actually have a higher space density 
(e.g. Phillipps \& Driver 1995; Driver et al., 1995b), then 
their clustering amplitude drops correspondingly. 
In any of these cases there is certainly no evidence for strong clustering 
of the faint blue galaxies about $L^\ast$ primaries: the numbers 
observed are consistent with (or more likely less than) the 
numbers expected for galaxies with `average' clustering ($r_{0}
\simeq 10$ Mpc; see e.g. Peebles 1980). 
The $1\sigma$ limit $0 < r_0 \la 4\,\mbox{Mpc}$ for the 
cross-correlation would be consistent with the low clustering amplitude 
seen for the faint blue galaxies themselves (see e.g. Efstathiou 
et al. 1991; Roukema \& Peterson 1994). 
In particular, Brainerd, Smail \& Mould (1995) find that their data 
at $\mbox{R} \sim 24$ ($\mbox{B} \sim 25$) are consistent with 
$r_{0} \simeq 2.0\,\mbox{Mpc}\,.$ 
Similar results are implied by the work of Couch, Jurcevic \& Boyle 
(1993) and Roche et al. (1993). Brainerd et al. conclude that 
the clustering evolution that would be needed if the faint blue galaxies 
(or rather, their present day descendents) had the same correlation 
function as `normal' giants is physically implausible, and prefer to 
identify them with a weakly clustered component (perhaps dwarf 
irregulars, see e.g., Thuan et al. 1991, Santiago \& 
da Costa 1990). 
This is certainly consistent with our finding of very weak clustering 
of the faint blue population about giants, too (for redshifts
around 0.2 to 0.4). Using HST data, Burkey et al. (1994) have found 
relatively few close companions around field galaxies at $z \simeq 0.5$ 
to $0.7\,.$ 
Recently, Le F\`{e}vre et al. (1996) have used redshift data to 
show that magnitude $17\fm5 \leq \mbox{I}_{AB} \leq 22\fm5$ 
galaxies at redshifts $0.2 \leq z \leq 0.5$ exhibit 
weak spatial clustering, consistent with these results. 

     The study of the clustering between the other samples of galaxies 
reveals no evidence of correlations within the estimated errors. 
It should be noted that the contamination of the samples of faint 
galaxies by stars (estimated in 
Table~5) should not affect the results of Table~7. The presence of 
a {\it random} population of images in any galaxy sample will not affect 
the overdensity in excess of a random distribution found within 60~arcsec 
of other galaxy. The only effect of star contamination will be to 
increase the errors present in the overdensities; as the total number 
of images in the samples -- including stars -- have been used in the 
error calculations of Section~7 and Table~6, these results are unaffected. 
 
      Studies of the angular two-point correlation function of galaxies 
at faint magnitudes find an amplitude smaller than that 
for brighter magnitudes (e.g. Efstathiou et al. 1991; 
Neuschaefer et al. 1991; Roche et al. 1993). 
Adopting an amplitude of the correlation function of $w \simeq 0.02$ 
at an angular separation of $30"$ for B $\simeq23\fm5$ to $25\fm0$ 
galaxies (see the review of Efstathiou 1995), 
and an angular dependence of $w(\theta) \propto \theta^{\;0.8},$ 
predicts an excess of 1.7\% faint galaxies about other faint 
galaxies in the $10"$ to $60"$ separation range. This compares 
with the observed {\it under}density of $(7.7 \pm 5.5)\;\%$ for the 
faint blue galaxies of the current study.

\section{Conclusions}

    No evidence is found, within the errors in the data, for an 
association between the faint blue galaxies and L$^\ast$ candidates 
for separations less than 1~arcmin, corresponding to scales of 
$\la$ few hundred~kpc. This is 
inconsistent with some merger models of galaxy evolution which 
predict that the faint blue galaxies have nearby parent giants with 
which they later merge; if this effect did occur significantly, 
it would have to be at a redshift $> 0.4\,.$ 
Similarly, although a complete understanding of galaxy mergers is 
still lacking, recent work on the hierarchical merging of haloes 
(e.g. Kauffmann \& White 1993; Lacey \& Cole 1993; Kauffmann, 
Guiderdoni \& White 1994; Cole et al. 1994) may suggest that, 
while giant galaxies 
could well be built up in this way, the merger rate may not be sufficient 
to explain the very large numbers of faint galaxies actually seen 
(pre-merger sub-units in this theory). Our current result may therefore 
be additional evidence in favour of  a fading scenario over merging 
as being the fate of the faint blue galaxy population, at least to 
the extent that the latter are seen at moderate redshifts, as is 
usually presumed.

\section*{Acknowledgements}
The authors would like to acknowledge the contribution of the 
other Hitchhiker team members, in particular Craig McKay, Nick Rees, 
Hugh Lang and Rhys Morris. Richard Ellis is thanked for constructive 
comments relating to data reduction, image detection and faint galaxies. 
We also thank Rogier Windhorst for discussions on faint galaxy clustering 
and the HST results. Gerry Gilmore is thanked for providing his Galactic 
star number counts modelling program. Some data reduction and analysis 
was performed using computing equipment and software provided by the 
Starlink project. 
The William Herschel Telescope is operated on the island of La Palma by
the Royal Greenwich Observatory in the Spanish Observatorio del
Roque de los Muchachos of the Instituto de Astrofisica de
Canarias. We thank the staff of the observatory for their help and 
cooperation. 
Finally we thank all the scheduled observers who allowed
Hitchhiker to operate in parallel with their observations.

\section{References}

\begin{list}{}{ \setlength{\leftmargin}{10mm} 
     \setlength{\itemindent}{-10mm}  \setlength{\labelwidth}{0mm} 
     \setlength{\itemsep}{0mm}  }
\item[ Arag\'{o}n-Salamanca A.,] Ellis R. S., Schwartzenberg J. M.,  
    Bergeron J. A., 1993, ApJ 421, 27 
\item[ Bahcall J. N.,] Soneira R. M., 1980, ApJS 44, 73 
\item[ Bell R. A.,] Gustafsson B., 1978, A\&AS 34, 229 
\item[ Bell R. A.,] Gustafsson B., 1989, MNRAS 236, 653 
\item[ Bernstein G. M.,] Tyson J. A., Brown W. R., Jarvis J. F., 1994, 
     ApJ 426, 516 
\item[ Bessell M. S., 1979,] PASP 91, 589 
\item[ Brainerd T. G.,] Smail I. R., Mould J. R., 1995, MN 275, 781 
\item[ Broadhurst T. J.,] Ellis R. S., Glazebrook K., 1992, 
    Nature 355, 55 
\item[ Broadhurst T. J.,] Ellis R. S., Shanks T., 1988,  
    MNRAS 235, 827 
\item[ Bruzual G., 1983,] ApJ 273, 105. 
\item[ Burkey J. M.,] Keel W. C., Windhorst R. A., Franklin B. E., 1994, 
    ApJ 429, L13 
\item[ Carlberg R. G., 1992,] ApJ 399, L31 
\item[ Carlberg R. G.,] Charlot S., 1992, ApJ 397, 5 
\item[ Casertano S.,] Ratnatunga K. U., Griffiths R. E., Im M., 
    Neuschaefer L. W., Ostrander E. J., Windhorst R. A., 1995, 
    ApJ 453, 599 
\item[ Cawson M. G. M., 1983,] Ph.D. thesis, University of Cambridge. 
\item[ Cole S.,] Arag\'{o}n-Salamanca A., Frenk C. S., Navarro J., 
    Zepf S. E., 1994, MNRAS 271, 781
\item[ Cole S.,] Ellis R. S., Broadhurst T. J., Colless M. M., 1994,  
    MNRAS 267, 541 
\item[ Coleman G. D.,] Wu C.-C., Weedman D. W., 1980, 
    ApJS 43, 393 
\item[ Colless M. M.,] Ellis R. S., Broadhurst T. J., K. Taylor K.,  
    Peterson B. A., 1993, MNRAS 261, 19 
\item[ Colless M. M.,] Ellis R. S., Taylor K., Hook R. N., 1990, 
    MNRAS 244, 408 
\item[ Colless M. M.,] Schade D., Broadhurst T. J., Ellis R. S., 
    1994, MNRAS  267, 1108 
\item[ Couch W. J.,] Jurcevic J. S., Boyle B. J., 1993, MNRAS, 260, 241
\item[ Cousins A. W. J.,] 1976, Mem. R. Astron. Soc. 81, 25 
\item[ Cowie L. L.,] Hu E. M., Songaila A., 1995, AJ 110, 1576 
\item[ Cowie L. L.,] Songaila A., Hu E. M., 1991, Nat 354, 460 
\item[ Dalcanton J. J., 1993,] ApJ 415, L87
\item[ Driver S. P.,] 1994, Ph.D. thesis, University of Wales, Cardiff. 
\item[ Driver S. P.,] Phillipps S., Davies J. I., Morgan I., Disney M. J., 
    1994a, MNRAS 266, 155 
\item[ Driver S. P.,] Phillipps S., Davies J. I., Morgan I., Disney M. J., 
    1994b, MNRAS 268, 393
\item[ Driver S. P.,] Windhorst R. A., Ostrander E. J., Keel W. C., 
    Griffiths R. E., Ratnatunga K. U., 1995a, ApJ 449, L23 
\item[ Driver S. P.,] Windhorst R. A., Griffiths R. E.,1995b, ApJ 453, 48 
\item[ Efstathiou G.,] 1995, MNRAS 272, L25 
\item[ Efstathiou G.,] Bernstein G., Katz N., Tyson J. A., Guhathakurta P., 
    1991, ApJ 380, L47 
\item[ Efstathiou G.,] Ellis R. S., Peterson B. A., 1988, MNRAS 232, 431 
\item[ Giraud E., 1992,] A\&A 257, 501 
\item[ Gilmore G., 1984,] MNRAS 207, 223 
\item[ Gilmore G.,] King I. R., van der Kruit, P. C., 1990, in 
    {\em The Milky Way as a Galaxy,} University Science Books, 
    Mill Valley, California 
\item[ Gilmore G.,] Wyse R. F. G., Kuijken K., 1989, ARA\&A 27, 555 
\item[ Glazebrook K.,] Ellis R. S., Colless M. M., Broadhurst T. J., 
    Allington-Smith J., Tanvir N., 1995a,  MNRAS 273, 157 
\item[ Glazebrook K.,] Ellis R. S., Santiago B., Griffiths R. E., 1995b, 
    MNRAS 275, L19 
\item[ Glazebrook K.,] Peacock J. A., Miller L., Collins C. A., 1995c,
    MNRAS 275, 169 
\item[ Jones L. R.,] Fong R., Shanks T., Ellis R. S., Peterson B. A., 
    1991, MNRAS 249, 481
\item[ Jones J. B.,] Driver S. P., Davies J. I., Phillipps S., Morgan I., 
    Disney M. J., 1994, In the Proceedings, I.A.U. Symposium No.~161, 
    {\it Astronomy from Wide-field Imaging,}  ed. H. T. MacGillivray,
    E. B. Thomson, B. M. Lasker, I. N. Reid, D. F. Malin, R. M. West \& 
    H. Lorenz, Kluwer Academic Publishers, pp. 77-78. 
\item[ Kauffmann G.,] Guiderdoni B., White S. D. M., 1994, MNRAS 267, 981
\item[ Kauffmann G.,] White S. D. M., 1993, MNRAS 261, 921
\item[ Koo D. C., 1990,] In the Proceedings, Edwin Hubble Centennial 
    Symposium, {\it Evolution of the Universe of Galaxies,} 
    ed. R. G. Kron, Astron. Soc. Pacific, San Francisco, pp. 268--285. 
\item[ Koo D. C.,] Guzm\'{a}n R., Faber S. M., Illingworth G. D., 
    Bershady M. A., Kron R. G., Takamiya M., 1995, ApJ 440, L49
\item[ Koo D. C.,] Kron R. G., 1992, ARA\&A 30, 613
\item[ Kron R. G.,] 1980, ApJS, 43, 305
\item[ Lacey C.,] Cole S., 1993, MNRAS 262, 627
\item[ Landolt A. U.,] 1983, AJ 88, 439 
\item[ Landy S. D.,] Szalay A., 1993, ApJ 412, 64
\item[ Le F\`{e}vre O.,] Hudon D., Lilly S. J., Crampton D., Hammer F., 
    Tresse L., 1996, ApJ, in press
\item[ Lilly S. J.,] Cowie L. L., Gardner J. P., 1991, ApJ 369, 79
\item[ Longair M. S.,]  Seldner M., 1979, MNRAS 189, 433
\item[ Metcalfe N.,] Shanks T., Fong R., Jones L. R., 1991, MNRAS 249, 498
\item[ Metcalfe N.,] Shanks T., Fong R., Roche N., 1995, MNRAS 273, 257
\item[ Neuschaefer L.,] Windhorst R. A., 1995, ApJ 439, 14
\item[ Neuschaefer L.,] Windhorst R. A., Dressler A., 1991, ApJ 382, 32 
\item[ Peebles P. J. E.,] 1980, {\it The Large-Scale Structure of the 
    Universe,} Princeton University Press, Princeton, New Jersey. 
\item[ Phillipps S., 1985a,] MNRAS 212, 657 
\item[ Phillipps S., 1985b,] Nature 314, 721 
\item[ Phillipps S.,] Driver S. P., 1995, MNRAS 274, 832 
\item[ Phillipps S.,] Shanks T., 1987a, MNRAS 227, 115 
\item[ Phillipps S.,] Shanks T., 1987b, MNRAS 229, 621 
\item[ Pritchet C. J.,] Infante L., 1992, ApJ 399, L35 
\item[ Rocca-Volmerange B.,] Guiderdoni B., 1990, MNRAS 247, 166
\item[ Roche N.,] Shanks T., Metcalfe N., Fong R., 1993, MNRAS 263, 360
\item[ Roukema B. F.,] Peterson B. A., 1994, A\&A 285, 361 
\item[ Santiago B. X.,] da Costa L. N., 1990, ApJ 362, 386 
\item[ Schechter P., 1976,] ApJ 203, 297 
\item[ Thuan T. X.,] Alimi J.-M., Gott J. R., Schneider S. E., 1991, 
    ApJ 370, 25 
\item[ T\'{o}th G.,] Ostriker J. P., 1992, ApJ 389, 5 
\item[ Turner J. A.,] Phillipps S., Davies J. I., Disney M. J., 1993, 
    MNRAS 261, 39 
\item[ Tyson J. A., 1988,] AJ 96, 1 
\item[ Woods D.,] Fahlman G. G., Richer H. B., 1995, ApJ 453, 583 
\item[ Yee H. K. C.,] Green R. F., 1987, ApJ 319, 28
\item[ Zepf S. E.,] Koo D. C., 1989, ApJ 337, 34 
\end{list}
\newpage
%
%  NOW FOR THE FIGURE CAPTIONS. 
%
%
\mbox{} \\[2mm]
\noindent {\bf \Large FIGURE CAPTIONS} \\[4mm]

\noindent {\bf Fig.~1.} Example reduced CCD frame. The 
R-band data frame is shown for the 
right ascension $12^{^h} 29.5^{^m},$ declination $+26^{^o} 26'$ 
(epoch 1950.0) field of Table~1. The candidate L$^\ast$ 
(large circles) and faint blue galaxies (small circles) from 
the samples of Section~5.1 are indicated.   \\[5mm]

\noindent {\bf Fig.~2.} A comparison of radii of variously defined 
photometric apertures as a function of total galaxy magnitude. 
The curves represent the sizes of circular apertures defined in 
six different ways for simulated face-on L$^\ast$ exponential disc 
galaxies. The locus for radii chosen to contain 90\% of the light 
is shown, as is that corresponding to detection at the 
26.0~mag.~(arcsec)$^{-2}$ isophote. Radii set at 2.5 times the 
Kron (1980) radius are presented. The results of Equation~(1) 
are given for indices $n= 1.0, 1.5$ and 2.0. An index $n= 1.5$ 
was selected for the photometry of Section~4. \\[5mm]

\noindent {\bf Fig.~3.} The (B--R)$_{_{\rm C}}$ -- B colour--magnitude 
diagram for the 
detected images of all fields, showing the candidate L$^\ast,$ the 
faint blue and the faint red galaxy samples. 
The solid curve is the locus corresponding to 
the no-evolution model of Sa-type giants. The dotted curve is the 
Sa giant locus displaced by an amount corresponding to Bruzual's (1983) 
$\mu = 0.5$ models. The dashed line illustrates the predicted 
completeness limit for the deepest field; all galaxies beyond this 
limit are rejected from the samples. \\[5mm]

\noindent {\bf Fig.~4.} The distribution of galaxy--galaxy separations 
between the  candidate L$^\ast$ and faint blue galaxies. Observational 
results are shown using a solid line. The dotted line represents 
randomly distributed simulated faint blue galaxies, normalised to the 
scale of the observational results. \\[5mm]
%
%
%
%
%
%
%
%
%
%   TABLE 1
% 
%
\begin{table*}[t]
\noindent {\bf Table 1.} Observational data \\
\begin{flushleft}
\begin{tabular}{cclcccc}
\cline{1-7} \\
R.A.  & Dec.  & 
   Date  & \multicolumn{2}{c}{ Integration time (min.) } &
   R band seeing & Area of reduced   \\
{\scriptsize (1950.0)} & {\scriptsize (1950.0)} & & \hspace{2mm} B band 
   \hspace{1mm} & \hspace{3mm} R band \hspace{1mm} &
   (FWHM) & data frame \\[3mm]
\cline{1-7}   \\[-1mm]
$12^{^h} 29.5^{^m}$  &  $+26^{^o} 26'$  &  1991 Feb 15 & 120 & 120 & 
   1.6 \mbox{arcsec} & $5.5' \times 3.0'$ \\[3mm]
 11~ 40.1~ & +19~ 49~~ & 1991 Feb 12        & 75 & 75 & 1.3~\hspace{8mm} & 
    $5.5 \times 3.0$ \\[3mm]
 13~ 37.8~ & +11~ 28~~ & 1993 Feb 28, & 40 & 60 & 1.6~\hspace{8mm} & 
    $5.0 \times 3.3$ \\
           &           & \hspace{6mm} Mar 2 & & &     & \\[3mm]  
 15~ 47.6~ & +21~ 35~~ & 1993 May 22        & 32 & 32 & 1.6~\hspace{8mm} & 
    $5.3 \times 3.1$  \\[3mm]
\cline{1-7}  \\[8mm]
\end{tabular}
\end{flushleft}
\end{table*}
%
%
%
%
%
%
%
%
%
%
%   TABLE 2
% 
\setcounter{table}{1}
\begin{table*}[t]
\noindent {\bf Table 2.} Observational results for the association of faint
blue galaxies with candidate L$^\ast$ galaxies  \\
\begin{flushleft}
\begin{tabular}{ccc} 
\cline{1-3} \\
Separation range & Observed number of galaxy-- & Predicted number 
   of separations \\
 (in arcsec) & galaxy separations & for random distributions of 
  blue galaxies \\[3mm]
\cline{1-3}   \\[-1mm]
    all   & 525 & $525.00\dagger$ \\[3mm]
 10 -- 60 & $92 \pm 10$ & ~92.26 \\[3mm]
 15 -- 60 & $88 \pm 9$  & ~88.40 \\[3mm]
\cline{1-3}  \\
\end{tabular}
\end{flushleft}
$\dagger$ Set by the normalisation to the number of observational 
results. \\[8mm]
\end{table*}
%
%
%
%
%
%
%
%
%
%
%
%   TABLE 3
%
\setcounter{table}{2}
\begin{table*}[t]
\noindent {\bf Table 3.} Statistics for the detection of faint galaxy 
images in the vicinity of four candidate L$^\ast$ galaxies  \\
\begin{flushleft}
\begin{tabular}{ccccc} 
\cline{1-5} \\
Separation between & Number of  & Number of times &
   Number of times & Number of times faint \\
faint galaxy and L$^\ast$  & simulations & faint and L$^\ast$ & faint galaxy 
 was & galaxy was merged  \\
candidate (arcsec)  & considered & galaxies merged & a distinct image & 
 with a third image  \\[3mm]
\cline{1-5}   \\
 ~5.0 & 32 & 32 & ~0 & ~0 \\[3mm]
 ~7.5 & 32 & 30 & ~2 & ~0 \\[3mm]
 10.0 & 32 & ~5 & 18 & ~9 \\[3mm]
 12.5 & 32 & ~0 & 20 & 12 \\[3mm]
 15.0 & 32 & ~0 & 22 & 10 \\[3mm]
\cline{1-5}  \\[8mm]
\end{tabular}
\end{flushleft}
\end{table*}
%
%
%
%
%
%
%
%
%
%
%   TABLE 4
% 
\setcounter{table}{3}
\begin{table*}
\noindent {\bf Table 4.} Selection criteria for samples of galaxies \\
\begin{flushleft}
\begin{tabular}{lccccc}
\cline{1-6} \\
Galaxy sample & Magnitude limits & Colour index limits & 
  Star/galaxy & Visual & Number in \\
  & & & test & test & sample \\[3mm]
\cline{1-6}   \\
L$^\ast$ candidates & $20{\fm}5 \leq \mbox{B} \leq 22{\fm}0$ & 
  See Section~5.1 & Yes & Yes & ~13 \\[3mm]
Faint blue & $23{\fm}5 \leq \mbox{B} \leq 25{\fm}0$ & $-0{\fm}4 \leq$ 
  (B--R)$_{_{\rm C}} \leq +1{\fm}2$  & No & No & 152 \\[3mm]
Bright & $20{\fm}5 \leq \mbox{B} \leq 22{\fm}0$ &  $-0{\fm}5 \leq$ 
  (B--R)$_{_{\rm C}} \leq +3{\fm}5$  & Yes & Yes & ~21 \\[3mm]
Faint red & $23{\fm}5 \leq \mbox{B} \leq 25{\fm}0$ & $+1{\fm}2 \leq$ 
  (B--R)$_{_{\rm C}} \leq +3{\fm}5$  & No & No & 138 \\[3mm]
All faint & $23{\fm}5 \leq \mbox{B} \leq 25{\fm}0$ & $-0{\fm}5 \leq$ 
  (B--R)$_{_{\rm C}} \leq +3{\fm}5$  & No & No & 304 \\[3mm]
\cline{1-6}  \\
\end{tabular}
\end{flushleft}
\begin{flushleft}
The selection criteria for the samples of galaxies are summarised in the 
table. An additional constraint was imposed that the image lay within the 
expected selection limits of the 1995 February 15 data. In practice 
this affected the most extreme faint blue galaxies only.    \\[8mm]
\end{flushleft}
\end{table*}
%
%
%
%
%
%
%
%   TABLE 5
% 
\setcounter{table}{4}
\begin{table*}
\noindent {\bf Table 5.} Star contamination of the galaxy samples \\
\begin{flushleft}
\begin{tabular}{ccccccc}
\cline{1-7} \\
 \multicolumn{2}{c}{Field}  &  \multicolumn{2}{c}{Galactic coordinates}   &  
      \multicolumn{3}{c}{Estimated fractional star contamination}  \\[3mm]
 R.A.  & Dec. & ~~Long. & Lat. & Faint blue sample & Faint red sample & 
    All faint sample  \\[3mm]
\cline{1-7}   \\
$12^{^h} 29\fm5$ & $\!\!\! +26^{^o} 26'$ & $224^{^o}$ & $+86^{^o}$  &
     $1.0\,\%$ & ~$9.4\,\%$ & ~$3.5\,\%$ \\[3mm]
 11~ 40.1~ & +19~ 49~~ & 235~ & +73~  &  $1.0\,\%$ & ~$8.6\,\%$ & 
     ~$3.2\,\%$ \\[3mm]
 13~ 37.8~ & +11~ 28~~ & 341~ & +70~  & $1.7\,\%$ & $18\,\%$   & 
     ~$6.4\,\%$ \\[3mm]
 15~ 47.6~ & +21~ 35~~ & ~35~ & +49~  & $2.7\,\%$ & $37\,\%$~  & 
     $13\,\%$~ \\[3mm]
\cline{1-7}  \\[8mm]
\end{tabular}
\end{flushleft}
\end{table*}
%
%
%
%
%
%
%
%
%   TABLE 6
% 
\setcounter{table}{5}
\begin{table*}
\noindent {\bf Table 6.} Galaxy--galaxy separation results for additional
samples of galaxies \\
\begin{flushleft}
\begin{tabular}{llcccc}
\cline{1-6} \\
 First sample & Second sample & \multicolumn{2}{c}{Number of observed 
  separations} & \multicolumn{2}{c}{Normalised number from 
  simulations} \\[3mm]
  & & total &  $10"$ to $60"$ & 
  total range & ~~~$10"$ to $60"$   \\[3mm]
\cline{1-6}   \\[-1mm]
~~Bright  & ~~Faint blue & 860 & $146 \pm 15$ & 860.0 & 144.95 \\[3mm]
~~Faint blue & ~~Faint red & 4753 & $847 \pm 36$ & 4753.0 & ~~887.02 (a) \\
            &           &      &               &         & ~~851.12 (b)
\\[3mm]
~~Bright & ~~All faint & 1586 & $289 \pm 24$ & 1586.0 & 279.25  \\[3mm]
~~Faint blue & ~~Faint blue & 6178 & $994 \pm 59$ & 6178.0 & ~~1087.83 (a) \\
            &            &      &               &         & ~~1076.72 (b) \\[3mm]
\cline{1-6}  \\[3mm]
\end{tabular}
\end{flushleft}
\mbox{ } \\[-10mm]
\begin{flushleft}
The numbers of galaxy--galaxy separations between the various samples 
are presented for the observational data for both the entire range of 
separations and for the restricted range $10"$ to $60"$. Error estimates 
are taken from Section~7. 
For comparison the results of the Monte Carlo simulations for randomly 
distributed points are presented in columns 5 and~6 for the same separation 
ranges. For the faint blue -- faint red study, simulated case (a) 
refers to random blue and the observed red galaxies; case~(b) 
refers to the observed blue and simulated red galaxies. 
In the faint blue -- faint blue study, case~(a) represents the 
statistics for one observed and one random blue sample, while case~(b) 
refers to the intercorrelations of a single random sample. \\[5mm]
\end{flushleft}
\end{table*}
%
%
%
%
%
%
%
%
%   TABLE 7
% 
\setcounter{table}{6}
\begin{table*}
%   \mbox{ } \\[0mm]
\noindent {\bf Table 7.} Statistical conclusions \\
\begin{flushleft}
\begin{tabular}{llcc}
\cline{1-4} \\
 First sample & Second sample & Observed overdensity & Overdensity / 
   error \\[3mm]
\cline{1-4}   \\
~~Candidate L$\ast$ & ~~Faint blue & $0.0 \pm 10$ & $-0.0$  \\[3mm]
~~Bright  & ~~Faint blue    & $+1 \pm 15$     & $+0.1$ \\[3mm]
~~Faint blue & ~~Faint red  & $-40 \pm 36$ (a)& $-1.1$  \\
             &              & ~$-4 \pm 36$ (b)& $-0.1$  \\[3mm]
~~Bright     & ~~All faint  & $+10 \pm 24$    & $+0.4$  \\[3mm]
~~Faint blue & ~~Faint blue & $-94 \pm 59$ (a)& $-1.5$  \\
             &              & $-83 \pm 59$ (b)& $-1.4$  \\[3mm]
\cline{1-4}  \\[3mm]
\end{tabular}
\end{flushleft}
\begin{flushleft}
The excess densities over random distributions of galaxies are 
presented for each pair of samples considered. 
Simulation sets (a) and (b) are as in Table~6 for the faint blue--faint 
red and faint blue--faint blue studies. \\[10mm]
\end{flushleft}
\end{table*}
\mbox{ } \\[5mm]
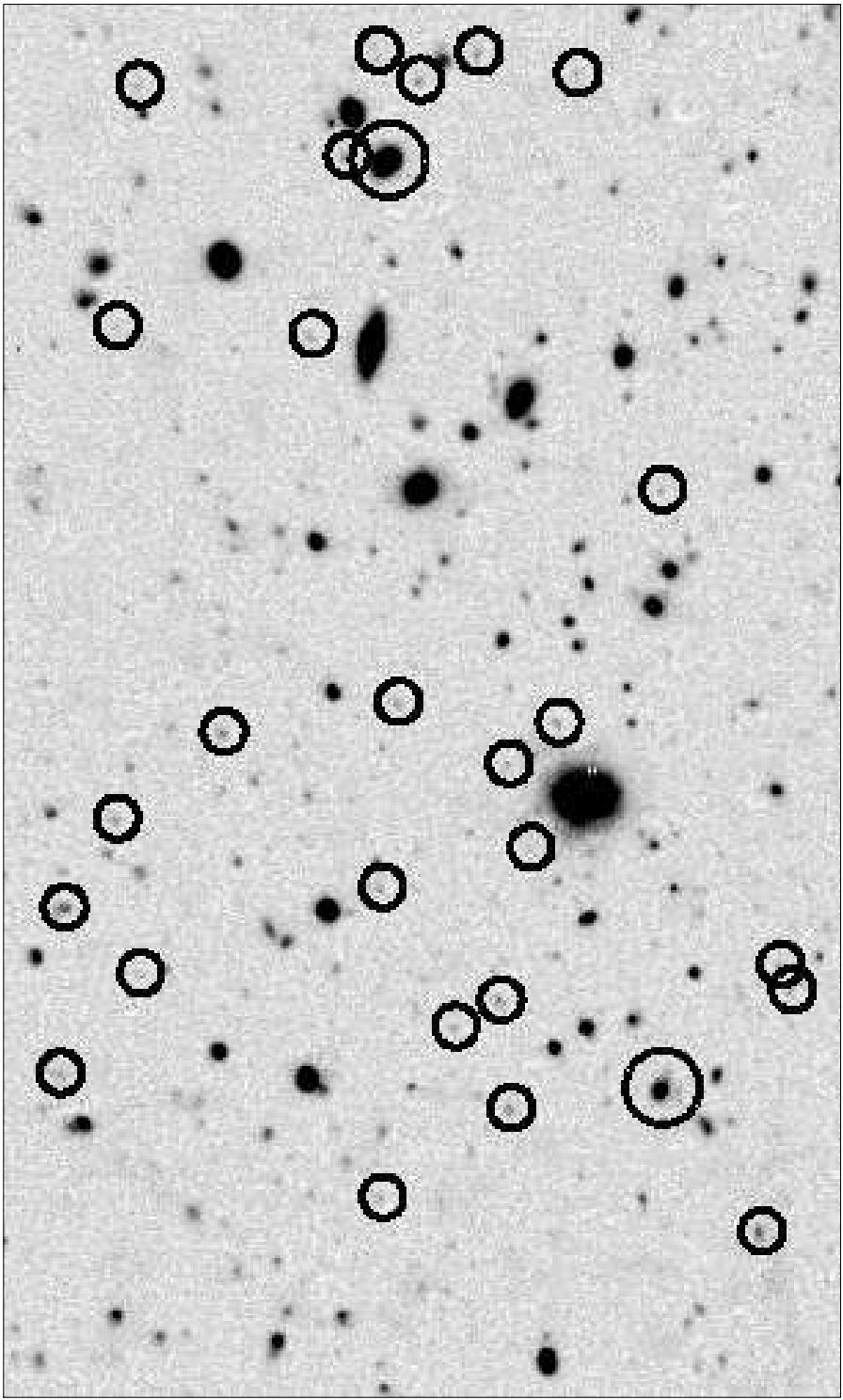
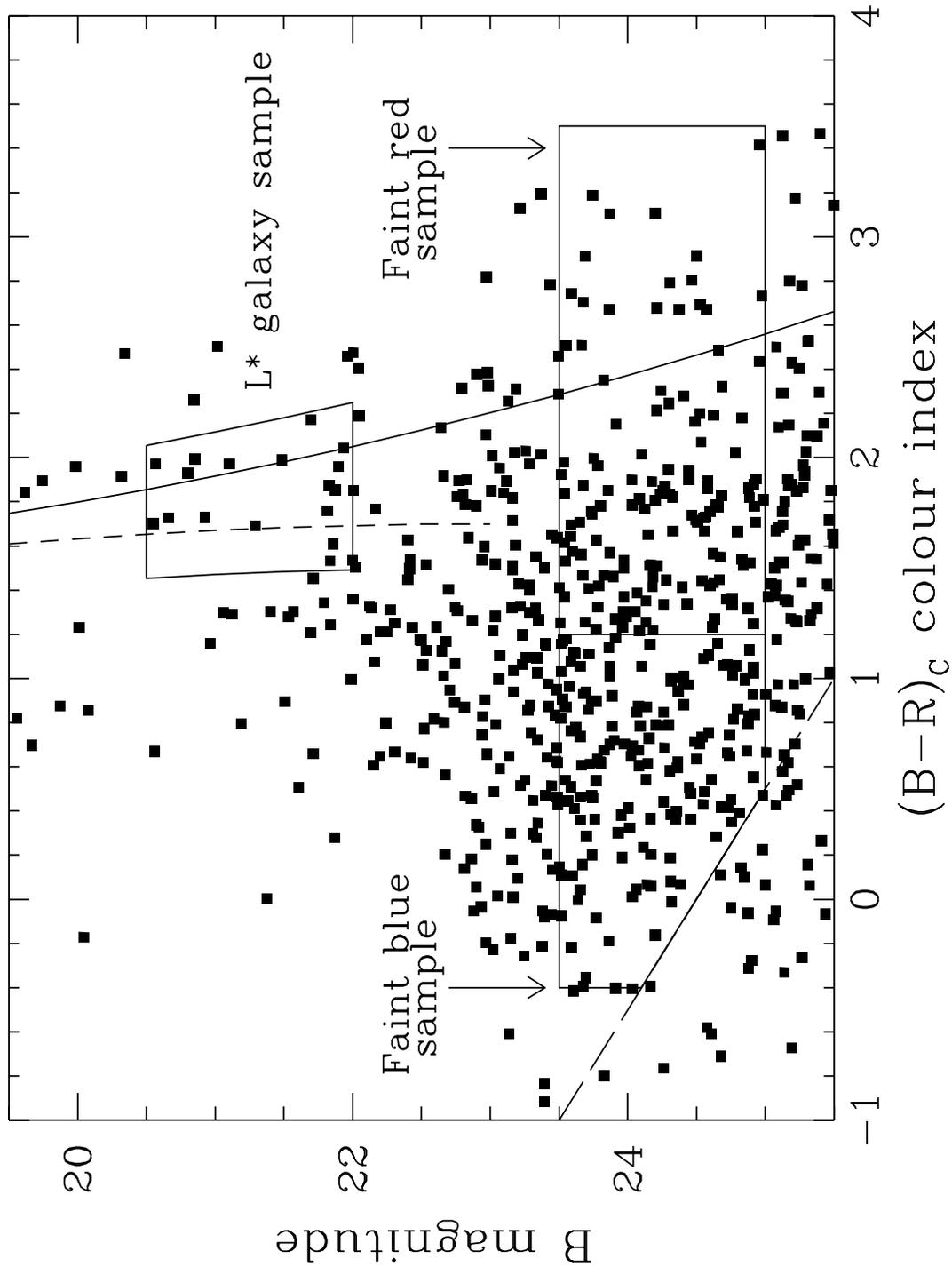
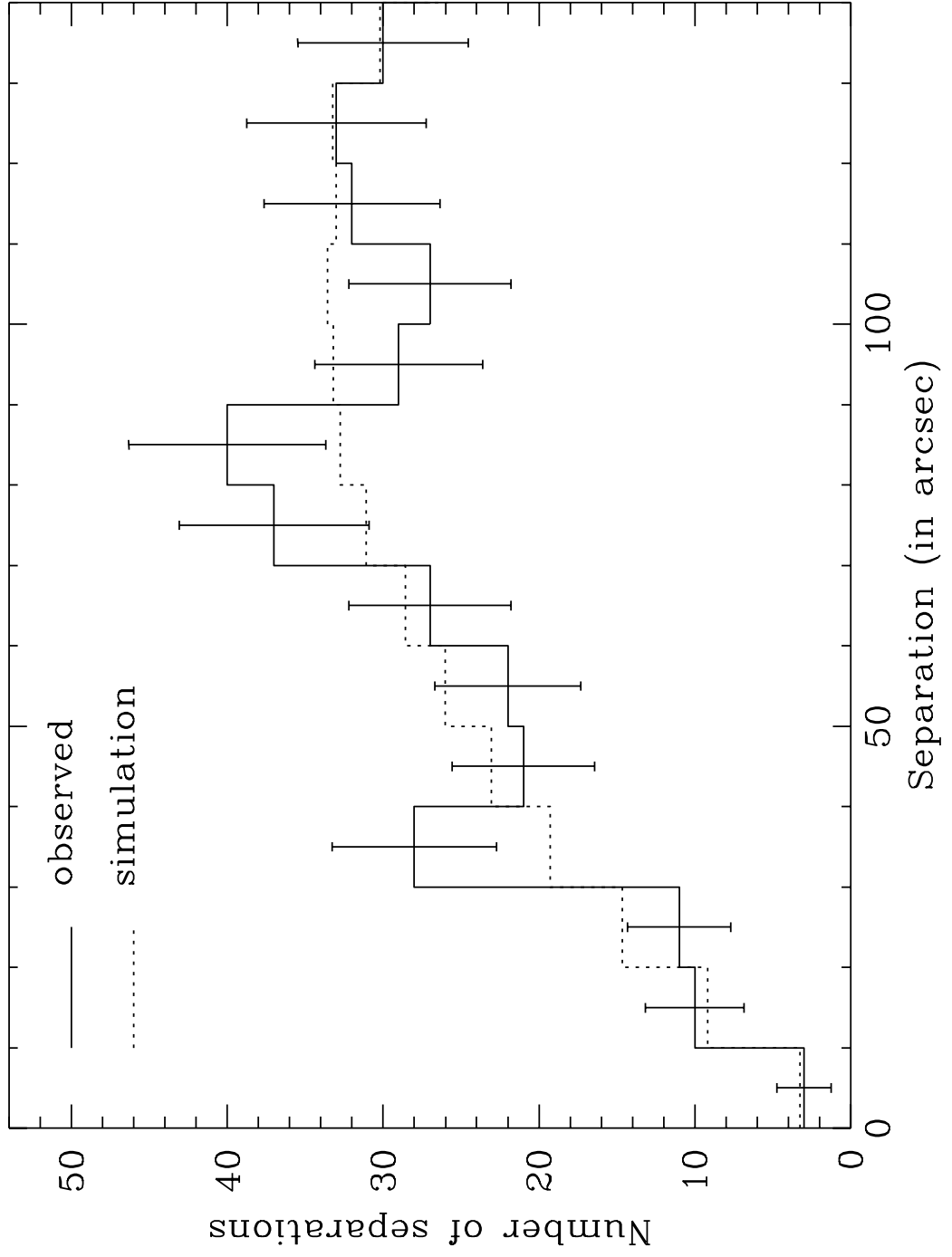
\end{document}